\documentclass[%
reprint,
showkeys,
nofootinbib,
amsmath,amssymb,
aps,
prd,
]{revtex4-2}
\usepackage[marginal]{footmisc}
\usepackage{graphicx}
\usepackage{dcolumn}
\usepackage{bm}
\usepackage[colorlinks,allcolors=black,CJKbookmarks=True]{hyperref}
\usepackage{color}

\usepackage{amsfonts}
\usepackage{mathrsfs}
\usepackage{amsmath}
\usepackage{amssymb}
\usepackage{xcolor}
\usepackage{mathtools}
\usepackage{tensor}
\usepackage{physics}
\usepackage{graphicx}
\usepackage{bm}
\usepackage{upgreek}
\usepackage{braket}
\usepackage{color,soul}
\usepackage{csquotes}
\usepackage{caption}
\usepackage{subcaption}
\usepackage{tabularx}
\usepackage{simplewick}
\usepackage[section]{placeins}
\usepackage{enumerate}
\usepackage{comment}
\usepackage{ulem}
\newcommand{\pp}{\uppi}

\newcommand{\dif}{\mathrm{d}}

\begin{document}

\title{Bispectrum of induced gravitational waves in the poltergeist mechanism}
	
	\author{Han-Wen Hu$^{1,2}$}
	\email{huhanwen@itp.ac.cn}
	
	\author{Cheng-Jun Fang$^{1,2}$}
	\email{fangchengjun@itp.ac.cn}

	\author{Zhen-Min Zeng$^{1,2}$}
	\email{cengzhenmin@itp.ac.cn}

        \author{Zong-Kuan Guo$^{1,2,3}$}
	\email{guozk@itp.ac.cn}

        \affiliation{$^{1}$Institute of Theoretical Physics, Chinese Academy of Sciences, Beijing 100190, China}
	\affiliation{$^{2}$School of Physical Sciences, University of Chinese Academy of Sciences, Beijing 100049, China }
    \affiliation{$^{3}$School of Fundamental Physics and Mathematical Sciences, Hangzhou Institute for Advanced Study, University of Chinese Academy of Sciences, Hangzhou 310024, China }

\begin{abstract}
In the poltergeist mechanism the enhancement of induced gravitational waves (GWs) occurs due to a sudden transition from an early matter-dominated era to the radiation-dominated era. In this work, we calculate the bispectrum of induced GWs from the poltergeist mechanism by adopting the sudden transition approximation. We find that the tensor bispectrum peaks either in the equilateral or squeezed configurations, depending on scales. Such a characteristic behavior enables us to distinguish it from that from other GW generation mechanisms.

\end{abstract}

\maketitle

\section{Introduction}
\label{sec:intr}

Gravitational waves (GWs) have emerged as a powerful tool for probing the Universe, providing insights into phenomena ranging from black hole mergers to the dynamics of the early Universe~\cite{LIGOScientific:2016aoc,LIGOScientific:2016dsl,LIGOScientific:2017ync}. Among the various cosmological sources of GWs, stochastic backgrounds induced by second-order curvature perturbations have raised a lot of interest in recent years~\cite{Ananda_2007,Baumann_2007,Bartolo_2007,Mangilli_2008,Assadullahi_2010,Kawasaki_2013,Dom_nech_2020,Cai_2019,Cai_2020,domenech_scalar_2021}. In general, there are two kinds of mechanism that can produce observable induced GWs. The first comes from the amplification of primordial curvature perturbations at small scales~\cite{Cai:2019bmk,Zhou_2020,Xu_2020,Di_2018,Mishra_2020,Fu_2020,balaji_induced_2022,Gao_2021}. This mechanism is usually accompanied by the formation of primordial black holes(PBHs). The other is the poltergeist mechanism~\cite{Inomata:2019ivs,Inomata:2019zqy,Inomata:2020lmk}, which occurs because of a rapid transition from a prolonged early matter-dominated (eMD) era to a radiation-dominated (RD) era.  This mechanism does not require an enhancement of primordial curvature perturbations at small scales. Instead, the amplification of induced GWs stems from the rapid oscillations of the scalar potential following the transition, particularly for modes that entered the horizon deep within the eMD era. These two mechanisms produce characteristically different GW signatures, which can be investigated by the GWs observatory, such as pulsar timing arrays (EPTA~\cite{EPTA:2023fyk} , PPTA~\cite{Reardon:2023gzh} , CPTA~\cite{Xu:2023wog} ,NANOGrav~\cite{NANOGrav:2023gor} , SKA~\cite{janssen2014gravitationalwaveastronomyska} ), space-based missions (LISA~\cite{amaroseoane2017laserinterferometerspaceantenna} , DECIGO~\cite{Kawamura:2011zz} , BBO~\cite{phinney2004big} , TaiJi~\cite{Hu:2017mde} , TianQin~\cite{TianQin:2015yph}) and ground-based Interferometer(aLIGO~\cite{LIGOScientific:2017ync}, Virgo~\cite{VIRGO:2014yos} , KAGRA~\cite{2019} , ET~\cite{Punturo:2010zz} , CE~\cite{Reitze:2019iox}).

In GW research, the power spectrum, which corresponds to the two-point correlation function of tensor perturbations, is commonly used to characterize the statistical properties of the signal. However, the three-point correlation function, or the gravitational wave bispectrum, becomes crucial when non-Gaussian features are present~\cite{gao2013bispectraprimordialscalartensor,Bartolo:2018rku,Peng:2024eok,Dimastrogiovanni_2019}. A nonzero bispectrum indicates the presence of nonlinear interactions during the generation or evolution of gravitational waves, which cannot be captured by the power spectrum alone. Moreover, the shape dependence of the bispectrum, such as local, equilateral, or folded configurations, provides a powerful diagnostic to distinguish between different generation mechanisms and test alternative theories of gravity. Therefore, studying the tensor bispectrum not only deepens our understanding of the early universe dynamics, but also offers observables for current and future gravitational wave detectors.

In particular, for second-order induced GWs, which are sourced from the quadratic terms of curvature perturbations, non-Gaussianity arises naturally, resulting in a nonvanishing bispectrum. Previous studies have investigated the tensor bispectrum generated by the first mechanism~\cite{Bartolo:2018rku}, where an enhancement of scalar perturbations on small scales leads to a corresponding enhancement in both the tensor power spectrum and the bispectrum.

In this work, we present the first calculation of the tensor bispectrum generated by the poltergeist mechanism. Adopting the sudden transition approximation throughout our analysis, we find that the bispectrum exhibits a critical scale-dependent behavior: its global maximum appears either in the equilateral or squeezed configuration, depending on the scale. This contrasts sharply with the case of purely RD scenarios, where the bispectrum peak invariably occurs in the equilateral configuration.

This paper is organized as follows. In Sec.~\ref{sec:model}, we briefly review the poltergeist mechanism, where we use the sudden transition approximation assumption. Based on this background, we analytically derive the general expression for the tensor bispectrum in this model in Sec.~\ref{sec:bis}, which provides the computational foundation for the numerical calculations of the bispectrum in Sec.~\ref{sec:num}. Section~\ref{se:Conclusion} presents a brief summary and discussion of our results.

\section{Model Setup}
\label{sec:model}

The scenario we consider in this paper is that the Universe experiences a rapid transition from a prolonged eMD era to a RD era before nucleosynthesis. Such an eMD epoch is motivated in a wide variety of contexts, including the decaying dark sector~\cite{Zhang_2015,hook_causal_2021,Erickcek_2021,Erickcek_2022,Nelson_2018,Blinov_2020}, PBHs~\cite{Sasaki:2018dmp,carr_constraints_2021,escriva_primordial_2023,escriva_primordial_2024}, post-inflationary reheating~\cite{Jedamzik_2010,Erickcek_2011,Fan_2014}, and solitons such as Q-balls~\cite{kusenko_supersymmetric_1998,Kasuya_2023}. Currently, there are no constraints on this epoch before nucleosynthesis. To obtain a relatively model-independent result, we adopt the sudden transition approximation commonly used in previous studies \cite{Inomata:2019ivs}, where the transition time is defined as $\eta_{\rm R}$. Before $\eta_{\rm R}$, the Universe is in the MD era and the equation of state parameter $\omega=0$ and after $\eta_{\rm R}$, the Universe goes back to the RD era ($\omega=1/3$) due to the rapid decay of matter. The scale factor $a(\eta)$ and the conformal Hubble parameter $\mathcal{H}(\eta)$ are given by
\begin{equation}
\frac{a(\eta)}{a\left(\eta_{\rm R}\right)}=\left\{\begin{aligned}
\left(\frac{\eta}{\eta_{\rm R}}\right)^2, \\
2 \frac{\eta}{\eta_{\rm R}}-1,
\end{aligned} \quad \quad \mathcal{H}(\eta)=\left\{ \begin{aligned}&\frac{2}{\eta}  \qquad\quad\quad\; \left(\eta \leq \eta_{\rm R}\right), \\
&\frac{1}{\eta-\eta_{\rm R} / 2}  \quad \left(\eta>\eta_{\rm R}\right).\end{aligned}\right.\right.
\end{equation}

If an early matter-dominated (eMD) era existed in the early Universe, it would significantly modify the evolution of scalar perturbations. For modes that entered the horizon during the MD phase, the Newtonian potential would remain nearly constant for an extended period after horizon reentry. When the Universe experiences the sudden transition back to the RD era, the Newtonian potential starts to oscillate rapidly, which leads to resonance amplification of induced gravitational waves.

We firstly review the generation mechanism of scalar-induced GWs in the poltergeist mechanism. For simplicity, we adopt the conformal Newtonian gauge and neglect the anisotropic stress in the energy-momentum tensor. In this gauge, the only independent scalar perturbation mode is the Newtonian potential $\Phi$, which serves as the source of scalar-induced GWs. The evolution equation for $k$-mode tensor perturbations $h_k$ then takes the form\cite{Kohri:2018awv}
\begin{equation}\label{eq:pert-eq-h}
h_{\boldsymbol{k}}^{\prime \prime}+2 \mathcal{H} h_{\boldsymbol{k}}^{\prime}+k^2 h_{\boldsymbol{k}}=\mathcal{S}(\boldsymbol{k}, \eta),
\end{equation}
where the source term $\mathcal{S}(\boldsymbol{k}, \eta)$ arises from quadratic interactions of scalar perturbations $\Phi$ and is explicitly given by
\begin{equation}\label{eq:source-term}
\begin{split}
    \mathcal{S}({\boldsymbol{k}},\eta)&=4  \int \frac{\dif^3 \boldsymbol{q}}{(2 \pp)^{3}} e (\boldsymbol{k}, \boldsymbol{q})\Bigg[2 \Phi_{\boldsymbol{q}} \Phi_{\boldsymbol{k}-\boldsymbol{q}}+\\
    &\frac{4}{3(1+\omega )}\left(\mathcal{H}^{-1} \Phi_{\boldsymbol{q}}^{\prime}+\Phi_{\boldsymbol{q}}\right)\left(\mathcal{H}^{-1} \Phi_{\boldsymbol{k}-\boldsymbol{q}}^{\prime}+\Phi_{\boldsymbol{k}-\boldsymbol{q}}\right)\Bigg],
\end{split}
\end{equation}
where $e(\boldsymbol{k}, \boldsymbol{q}) \equiv e^{i j}(\boldsymbol{k}, \boldsymbol{q}) q_i q_j $.
A standard approach to solving inhomogeneous differential equations in Eq.~\eqref{eq:pert-eq-h} is the Green function method, where the solution can be expressed as
\begin{equation}\label{eq:solution-h-k}
h_{\boldsymbol{k}}(\eta)=\frac{1}{a(\eta) }\int_{\eta_0}^\eta \dif \bar{\eta} G_{\boldsymbol{k}}(\eta ; \bar{\eta}) a(\bar{\eta}) \mathcal{S}(\boldsymbol{k}, \bar{\eta}),
\end{equation}
where the Green function $G_{\boldsymbol{k}}(\eta ; \bar{\eta})$ satisfies
\begin{equation}
G_{\boldsymbol{k}}^{\prime \prime}(\eta ; \bar{\eta})+\left(k^2-\frac{a^{\prime \prime}(\eta)}{a(\eta)}\right) G_{\boldsymbol{k}}(\eta ; \bar{\eta})=\delta(\eta-\bar{\eta}).
\end{equation}

 The scalar perturbation $\Phi$ can be expressed as $\Phi_{\boldsymbol{k}}(\eta) \equiv \Phi(k \eta) \phi_{\boldsymbol{k}}$, where $\Phi(k \eta)$ is the transfer function and $\phi_k$ is the initial value of the perturbation, which is generated from inflation.

 Then the source term Eq.\ \eqref{eq:source-term} can be expressed as
\begin{equation}\label{eq:source}
\mathcal{S}(\boldsymbol{k}, \eta)=\int \frac{\dif^3 \boldsymbol{q}}{(2 \pp)^3} e(\boldsymbol{k}, \boldsymbol{q}) f(\boldsymbol{k}, \boldsymbol{q}, \eta) \phi_{\boldsymbol{q}} \phi_{\boldsymbol{k}-\boldsymbol{q}},
\end{equation}
where
\begin{equation}
\begin{split}
    f(\boldsymbol{k}, \boldsymbol{q}, \eta)=&8 \Phi(q \eta) \Phi(|\boldsymbol{k}-\boldsymbol{q}| \eta)+\frac{16}{3(1+\omega)}\left[\mathcal{H}^{-1} \Phi^{\prime}(q \eta)+\Phi(q \eta)\right]\\
    &\times\left[\mathcal{H}^{-1} \Phi^{\prime}(|\boldsymbol{k}-\boldsymbol{q}| \eta)+\Phi(|\boldsymbol{k}-\boldsymbol{q}| \eta)\right],
\end{split}
\end{equation}
where $\omega$ is the equation of state parameter. Substituting Eq.~\eqref{eq:source} into Eq.~\eqref{eq:solution-h-k} and taking $x\equiv k \eta$, we obtain the following.
\begin{equation}
h_{\boldsymbol{k}}(\eta)=\int \frac{\dif^3 \boldsymbol{q}}{(2 \pp)^3} e(\boldsymbol{k}, \boldsymbol{q}) \phi_{\boldsymbol{q}} \phi_{\boldsymbol{k}-\boldsymbol{q}} \int_{0}^x \dif \bar{x} k G_{\boldsymbol{k}}(\eta ; \bar{\eta}) \frac{a(\bar{\eta})}{a(\eta)}  f(\boldsymbol{k}, \boldsymbol{q}, \bar{\eta}).
\end{equation}

\section{Theoretical Formulation of the Tensor Bispectrum}
\label{sec:bis}

The definition of the three-point correlation function is\cite{Bartolo:2018rku,Espinosa:2018eve}
\begin{equation}\label{eq:3-points-cor}
\begin{aligned}
    \left\langle h^r_{\boldsymbol{k}_1} h^s_{\boldsymbol{k}_2} h^t_{ \boldsymbol{k}_3}\right\rangle=&\int \frac{\dif^3 \boldsymbol{p}_1}{(2 \pp)^3}\frac{\dif^3 \boldsymbol{p}_2}{(2 \pp)^3}\frac{\dif^3 \boldsymbol{p}_3}{(2 \pp)^3} e^r\left(\boldsymbol{k}_1, \boldsymbol{p}_1\right)
 e^s\left(\boldsymbol{k}_2, \boldsymbol{p}_2\right)
 e^t\left(\boldsymbol{k}_3, \boldsymbol{p}_3\right)\\
    &\times \left\langle\phi_{\boldsymbol{p}_1}\phi_{\boldsymbol{k}_1-\boldsymbol{p}_1}\phi_{\boldsymbol{p}_2}\phi_{\boldsymbol{k}_2-\boldsymbol{p}_2}\phi_{\boldsymbol{p}_3}\phi_{\boldsymbol{k}_3-\boldsymbol{p}_3}\right\rangle \\&
    \times\int_{0}^x \dif x_1 k_1 G_{\boldsymbol{k}_1}(\eta ; \eta_1) \frac{a(\eta_1)}{a(\eta)}  f(\boldsymbol{k}_1, \boldsymbol{p}_1, \eta_1)\\
    &\times \int_{0}^x \dif x_2 k_2 G_{\boldsymbol{k}_2}(\eta ; \eta_2) \frac{a(\eta_2)}{a(\eta)}  f(\boldsymbol{k}_2, \boldsymbol{p}_2, \eta_2)\\
    &\times\int_{0}^x \dif x_3 k_3 G_{\boldsymbol{k}_3}(\eta ; \eta_3) \frac{a(\eta_3)}{a(\eta)}  f(\boldsymbol{k}_3, \boldsymbol{p}_3, \eta_3),
\end{aligned}
\end{equation}
where, $r,\ s,\ t$ denote polarization indices that take values
$(+)$ (plus polarization) and $\times$ (cross-polarization). We implicitly assume that the primordial curvature perturbations follow a Gaussian distribution. According to Wick's theorem, there are eight distinct nontrivial ways to contract the field operators in the above expression, and each contraction contributes equally to the correlation function. The two-point correlation of Newtonian potential is related to the primordial curvature perturbation spectrum \cite{Mukhanov:2005sc},
\begin{equation}
\left\langle\phi_{\boldsymbol{k}} \phi_{\boldsymbol{q}}\right\rangle\equiv \delta^3(\boldsymbol{k}+\boldsymbol{q}) \frac{2 \pp^2}{k^3}\left(\frac{3+3 \omega}{5+3 \omega}\right)^2 P_\zeta(k),
\end{equation}
where $P_{\zeta}(k)$ denotes the scalar perturbation power spectrum. A possible way to contract the six perturbation fields is
\begin{equation}
    \begin{aligned}
    &\contraction[2ex]{}{\phi}{_{\boldsymbol{p}_1}\phi_{\boldsymbol{k}_1-\boldsymbol{p}_1}
    \phi_{\boldsymbol{p}_2}\phi_{\boldsymbol{k}_2-\boldsymbol{p}_2}
    \phi_{\boldsymbol{p}_3}}{\phi}
    \phi_{\boldsymbol{p}_1}
    \contraction{}{\phi}{_{\boldsymbol{k}_1-\boldsymbol{p}_1}}{\phi}
    \phi_{\boldsymbol{k}_1-\boldsymbol{p}_1}
    \phi_{\boldsymbol{p}_2}
    \contraction{}{\phi}{_{\boldsymbol{k}_2-\boldsymbol{p}_2}}{\phi}
    \phi_{\boldsymbol{k}_2-\boldsymbol{p}_2}
    \phi_{\boldsymbol{p}_3}\phi_{\boldsymbol{k}_3-\boldsymbol{p}_3}\\
    &= (2 \pp)^9 \delta^3(\boldsymbol{k}_1+\boldsymbol{k}_2+\boldsymbol{k}_3) (2 \pp^2)^3 \left(\frac{3+3 \omega}{5+3 \omega}\right)^6 \\
    &\times\frac{P_\zeta(p_1)}{p_1^3} \frac{P_\zeta(p_2)}{p_2^3} \frac{P_\zeta(p_3)}{p_3^3} \delta^3(\boldsymbol{p}_1+\boldsymbol{k}_3-\boldsymbol{p}_3)\delta^3(\boldsymbol{k}_1-\boldsymbol{p}_1+\boldsymbol{p}_2).
    \end{aligned}
\end{equation}
Substituting the above results into Eq.~\eqref{eq:3-points-cor} yields
    \begin{equation}
\begin{aligned}
    &\left\langle h^r_{\boldsymbol{k}_1} h^s_{\boldsymbol{k}_2} h^t_{ \boldsymbol{k}_3}\right\rangle=(4 \pp^2)^3 \delta^3(\boldsymbol{k}_1+\boldsymbol{k}_2+\boldsymbol{k}_3)\left(\frac{3+3 \omega}{5+3 \omega}\right)^6 \\
    &\qquad\times\int \dif^3 \boldsymbol{p}_1\  e^r\left(\boldsymbol{k}_1, \boldsymbol{p}_1\right) e^s\left(\boldsymbol{k}_2, \boldsymbol{p}_2\right) e^t\left(\boldsymbol{k}_3, \boldsymbol{p}_3\right)\\
    &\qquad\times \frac{P_\zeta(p_1)}{p_1^3} \frac{P_\zeta(p_2)}{p_2^3} \frac{P_\zeta(p_3)}{p_3^3} \\
    &\qquad\times\int_{0}^x \dif x_1\ k_1 G_{\boldsymbol{k}_1}(\eta ; \eta_1) \frac{a(\eta_1)}{a(\eta)}  f(\boldsymbol{k}_1, \boldsymbol{p}_1, \eta_1)\\
    &\qquad\times \int_{0}^x \dif x_2\ k_2 G_{\boldsymbol{k}_2}(\eta ; \eta_2) \frac{a(\eta_2)}{a(\eta)}  f(\boldsymbol{k}_2, \boldsymbol{p}_2, \eta_2) \\
    &\qquad\times \int_{0}^x \dif x_3\ k_3 G_{\boldsymbol{k}_3}(\eta ; \eta_3) \frac{a(\eta_3)}{a(\eta)}  f(\boldsymbol{k}_3, \boldsymbol{p}_3, \eta_3).
\end{aligned}
\end{equation}
Although we retain the notation $p_2$ and $p_3$ in the above expression, they are actually defined by $\boldsymbol{p}_2=\boldsymbol{p}_1-\boldsymbol{k}_1$ and $ \boldsymbol{p}_3=\boldsymbol{p}_1+\boldsymbol{k}_3$. The third polarization tensor can be simplified as $e^t(\boldsymbol{k}_3, \boldsymbol{p}_3) = e^t(\boldsymbol{k}_3, \boldsymbol{p}_1+\boldsymbol{k}_3)=e^t(\boldsymbol{k}_3, \boldsymbol{p}_1)$. We now define the bispectrum as
\begin{equation}
    \left\langle h^r_{\boldsymbol{k}_1} h^s_{\boldsymbol{k}_2} h^t_{ \boldsymbol{k}_3}\right\rangle\equiv (2\pp)^3\delta^3(\boldsymbol{k}_1+\boldsymbol{k}_2+\boldsymbol{k}_3)\left\langle\!\left\langle h^r_{\boldsymbol{k}_1} h^s_{\boldsymbol{k}_2} h^t_{ \boldsymbol{k}_3}\right\rangle\!\right\rangle.
\end{equation}
It turns out that it is convenient to introduce the following variable:
\begin{equation}
\begin{split}
   &k_i u_i \equiv |\boldsymbol{k}_i-\boldsymbol{p}_i|, \quad k_i v_i \equiv p_i, \quad\\
   &\left(\frac{3+3 \omega}{5+3 \omega}\right)^2 f\left(\boldsymbol{k}, \boldsymbol{p}, \eta\right) \equiv 4 F\left(\boldsymbol{k}, \boldsymbol{p}, \eta\right).
\end{split}
\end{equation}
We define a kernel function which involves the time integral as
\begin{equation}
I(u, v, x)=\int_0^x \dif \bar{x} \frac{a\left(\bar{\eta}\right)}{a(\eta)} k G_{\boldsymbol{k}}\left(\eta, \bar{\eta}\right) F\left(u, v, \bar{\eta}\right),
\end{equation}
then we obtain
\begin{equation}\label{eq:bis-MD-RD}
    \begin{aligned}
    &\left\langle\!\left\langle h^r_{\boldsymbol{k}_1} h^s_{\boldsymbol{k}_2} h^t_{ \boldsymbol{k}_3}\right\rangle\!\right\rangle=(8 \pp)^3 \int \dif^3 \boldsymbol{p}_1\  e^r\left(\boldsymbol{k}_1, \boldsymbol{p}_1\right) e^s\left(\boldsymbol{k}_2, \boldsymbol{p}_2\right) e^t\left(\boldsymbol{k}_3, \boldsymbol{p}_1\right) \\
    &\qquad\times\frac{P_\zeta(p_1)}{p_1^3} \frac{P_\zeta(p_2)}{p_2^3} \frac{P_\zeta(p_3)}{p_3^3}\\
    &\qquad\times  I(u_1, v_1, x)I(u_2, v_2, x)I(u_3, v_3, x),
\end{aligned}
\end{equation}
for modes that reenter the horizon during the eMD era, the kernel function can be written as
\begin{equation}\label{eq:I-MD-RD}
\begin{aligned}
I\left(u, v, x\right)=&  \int_0^{x_{\rm R}} \dif \bar{x} \left(\frac{1}{2\left(x / x_{\rm R}\right)-1}\right)\left(\frac{\bar{x}}{x_{\rm R}}\right)^2\\
&\times k G_k^{\mathrm{MD} \rightarrow \mathrm{RD}}(\eta; \eta_1) F\left(u, v, \bar{x}\right) \\
&+\int_{x_{\rm R}}^x \dif \bar{x}\left(\frac{2\left(\bar{x} / x_{\rm R}\right)-1}{2\left(x / x_{\rm R}\right)-1}\right) k G_k^{\mathrm{RD}}(\eta; \eta_1) F\left(u, v, \bar{x}\right) \\
&\equiv  I_{\mathrm{MD}}\left(u, v, x\right)+I_{\mathrm{RD}}\left(u, v, x\right),
\end{aligned}
\end{equation}
where $I_{\rm MD}$ and $I_{\rm RD}$ denote the contribution during the eMD era and the radiation-dominated era, respectively.
The expressions of Green's functions are given by
\begin{equation}
k G_k^{\mathrm{RD}}(\eta, \bar{\eta})=\sin (x-\bar{x}),
\end{equation}
and
\begin{equation}
k G_k^{\rm MD \rightarrow RD}(\eta, \bar{\eta})=C\left(x, x_{\rm R}\right) \bar{x} j_1(\bar{x})+D\left(x, x_{\rm R}\right) \bar{x} y_1(\bar{x}),
\end{equation}
 where $j_1,\ y_1$ denote the 1st order spherical Bessel functions. Using the continuity of Green's function and its first derivative at $x=x_{\rm R}$, we can derive the coefficients:
\begin{equation}
C\left(x, x_{\rm R}\right)=\frac{\sin x-2 x_{\rm R}\left(\cos x+x_{\rm R} \sin x\right)+\sin \left(x-2 x_{\rm R}\right)}{2 x_{\rm R}^2},
\end{equation}
and
\begin{equation}
D\left(x, x_{\rm R}\right)=\frac{\left(2 x_{\rm R}^2-1\right) \cos x-2 x_{\rm R} \sin x+\cos \left(x-2 x_{\rm R}\right)}{2 x_{\rm R}^2}.
\end{equation}

The redefined source term $F(u,v,\bar{x})$ is given by
\begin{equation}\label{eq:source-MD-RD}
\begin{aligned}
    F\left(u, v, \bar{x}\right)=&\frac{3}{25(1+\omega)}\bigg[2(5+3 \omega) \Phi(u \bar{x}) \Phi(v \bar{x})\\
    &+4 \mathcal{H}^{-1}\left(\Phi^{\prime}(u \bar{x}) \Phi(v \bar{x})+\Phi(u \bar{x}) \Phi^{\prime}(v \bar{x})\right)\\
    &+4 \mathcal{H}^{-2} \Phi^{\prime}(u \bar{x}) \Phi^{\prime}(v \bar{x})\bigg],
\end{aligned}
\end{equation}
where $^{\prime}$ denotes the derivative with respect to the conformal time $\eta$. The evolution equation for the Bardeen potential is\cite{Mukhanov:2005sc}
\begin{equation}
\Phi^{\prime \prime}+3(1+\omega) \mathcal{H} \Phi^{\prime}+\omega k^2 \Phi=0 .
\end{equation}
Solving this equation, we obtain the piecewise solution
\begin{equation}\label{eq:potential-MD-RD}
\Phi\left(x, x_{\rm R}\right)= \left\{
\begin{aligned}
&1  \qquad\qquad\qquad\qquad\qquad\quad\;\;\;\, x < x_{\rm R}, \\
& A\left(x_{\rm R}\right) \mathcal{J}(x)+B\left(x_{\rm R}\right) \mathcal{Y}(x) \quad x \geq x_{\rm R},\end{aligned}
\right.
\end{equation}
where $\mathcal{J}(x)$ and $\mathcal{Y}(x)$ are constructed from the $j_1(x)$ and $y_1(x)$
\begin{equation}
\mathcal{J}(x)=\frac{3 \sqrt{3} j_1\left(\frac{x-x_{\rm R} / 2}{\sqrt{3}}\right)}{x-x_{\rm R} / 2},\quad \mathcal{Y}(x)=\frac{3 \sqrt{3} y_1\left(\frac{x-x_{\rm R} / 2}{\sqrt{3}}\right)}{x-x_{\rm R} / 2},
\end{equation}
meanwhile, the constant coefficients $A(x_{\rm R}),\ B(x_{\rm R})$ are determined by the continuity conditions of $\Phi(x)$ at $x=x_{\rm R}$,
\begin{equation}
\begin{split}
    &A\left(x_{\rm R}\right)=\frac{1}{\mathcal{J}\left(x_{\rm R}\right)-\frac{\mathcal{Y}\left(x_{\rm R}\right)}{\mathcal{Y}^{\prime}\left(x_{\rm R}\right)} \mathcal{J}^{\prime}\left(x_{\rm R}\right)},\\
    &B\left(x_{\rm R}\right)=-\frac{\mathcal{J}^{\prime}\left(x_{\rm R}\right)}{\mathcal{Y}^{\prime}\left(x_{\rm R}\right)} A\left(x_{\rm R}\right) .
\end{split}
\end{equation}

In order to calculate the bispectrum, we need to conduct a more careful analysis of $I(u,v,x)$. The $I_{\mathrm{MD}}$ component in Eq.~\eqref{eq:I-MD-RD} can be readily evaluated, while numerically computing the radiation-dominated era Green's function time integral $I_{\mathrm{RD}}$ requires a subtle approach. Note that the source term $F\left(u, v, \bar{x}\right)$ contains terms proportional to $\mathcal{H}^{-1}$ and $\mathcal{H}^{-2}$. To cast $I_{\mathrm{RD}}$ into the standard form for the radiation-dominated universe, notice that $\dif \eta \propto \dif a$, one can obtain that
\begin{equation}
    \frac{a^{\prime}(\eta)}{a_0}=\frac{1}{\eta_{0}},
\end{equation}
where $a_0\equiv a(\eta_0)$, $\eta_0$ is an arbitrary reference time, which implies
\begin{equation}
    \mathcal{H}(\eta)=\frac{1}{\eta-\eta_{\rm R}+\mathcal{H}^{-1}(\eta_{\rm R})} \equiv \frac{1}{\eta^{\prime}}.
\end{equation}
$\eta^{\prime}$ is the new "conformal time" variable. Leaving $x^{\prime}=k \eta^{\prime}$, the integral $I_{\rm RD}$ can be rewritten as
\begin{equation}
    I_{\rm RD}=\int_{x^{\prime}_0}^{x^{\prime}}\dif \bar{x}^{\prime}\frac{\bar{x}^{\prime}}{x^{\prime}} k G_k^{\mathrm{RD}}(\eta^{\prime}; \bar{\eta}^{\prime}) F\left(u, v, \bar{x}^{\prime}\right),
\end{equation}
where $x_{0}^{\prime}=k \eta^{\prime}(\eta_{\rm R})=k \mathcal{H}^{-1}(\eta_{\rm R})$. After this transformation, the lower limit of the time integral depends on the wavenumber $k$ of the mode. Furthermore, it should be noted that the coefficients $A\left(x_{\rm R}\right)$ and $B\left(x_{\rm R}\right)$ in the Bardeen potential are independent of coordinate transformations. Therefore, during numerical computations, their values derived from the original conformal time $\eta$ description can be directly adopted.

\section{Numerical Results and Configuration Analysis}
\label{sec:num}

We now proceed to compute the bispectrum of this transition model using numerical methods. We assume that the primordial scalar perturbation power spectrum takes a power-law form
\begin{equation}\label{eq:Ph-zeta-scalar}
P_\zeta(k)=\Theta\left(k_{\max }-k\right) A_{\zeta}\left(\frac{k}{k_*}\right)^{n_s-1},
\end{equation}
where $A_{\zeta} \simeq 2.1 \times 10^{-9}$ being the amplitude at the pivot scale, $n_{s}\simeq 0.97$ the spectral tilt, and $k_*=0.05\mathrm{Mpc^{-1}}$ the pivot scale~\cite{plank2020}.  The ultraviolet cut-off makes sure that the modes in which we are interested are all in the linear region, it provides a physical cut off for the integrated momentum $p_i\ (i=1,2,3)$. For modes that are larger than $k_{\max}$, the density contrast of matter $\delta_{m}$ will grow to a non-linear region and lead to the breakdown of linear perturbation theory. This cutoff condition can be estimated by $\delta_{m, k_{\max}} \sim 1$
\begin{equation}
    k_{\max}\simeq \frac{\eta_{\rm R}}{450}.
\end{equation}

Since we are interested in the enhanced mode that reenters the horizon in MD era, we need to introduce the lower bound of external momentum modes $k_{\min}$, which is defined by $k_{\min }=a\left(\eta_{\rm R}\right) H\left(\eta_{\rm R}\right)=\mathcal{H}\left(\eta_{\rm R}\right)$. Combining Eqs.~\eqref{eq:bis-MD-RD}, \eqref{eq:I-MD-RD}, \eqref{eq:source-MD-RD} and \eqref{eq:potential-MD-RD}, we can numerically calculate the dependence of the bispectrum $\left\langle\!\left\langle h_{k_1}^r h_{k_2}^s h_{k_3}^t\right\rangle\!\right\rangle$ on the external momentum $k_i\ (i=1,2,3)$ in various cases.

\begin{figure}[!htb]
    \centering
    \includegraphics[width=1\linewidth]{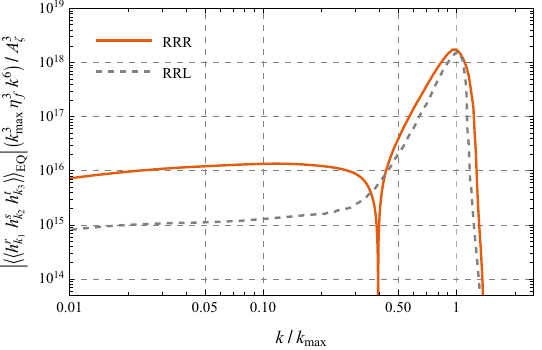}
    \caption{Bispectrum of gravitational waves in the matter-radiation transition model for the equilateral shape. We plot the dimensionless bispectra for two independent polarization modes (RRR, RRL) in the equilateral shape, represented by the red solid line and gray dashed line in the figure, respectively.}
    \label{fig:EQ-shape}
\end{figure}

First, for the sake of simplicity, we plot the bispectrum of gravitational waves with two different polarizations in the equilateral shape as shown in Fig.~\ref{fig:EQ-shape} to show their scale dependence. Noting that the bispectrum in the equilateral configuration exhibits the chiral symmetry of $\rm R\leftrightarrow L$, there are only two independent polarization modes\cite{Bartolo:2018rku}, RRR and RRL. When $k$ is small, the bispectrum of the equilateral shape for the RRR polarization is negative. As the scale decreases, the curve shows a zero point near $k\sim 0.39k_{\max}$. Subsequently, the RRR bispectrum increases with $k$, reaching a peak at $k\sim k_{\max}$, and then rapidly decreases, which is consistent with the case of the two-point correlation function\cite{Inomata:2019ivs}. The rapid decrease at $k>k_{\max}$ is also affected by the cutoff in the scalar perturbation power spectrum. The overall trend of the RRL polarization mode is the same as that of the RRR polarization, but there is no positive-negative alternation, but it remains negative throughout. Moreover, compared to the RRR term, the bispectrum of RRL polarization is suppressed, which is consistent with the bispectrum induced by the enhanced primordial curvature power spectrum in the pure RD era\cite{Bartolo:2018rku}.

\begin{figure}[!htb]
    \centering
    \includegraphics[width=1\linewidth]{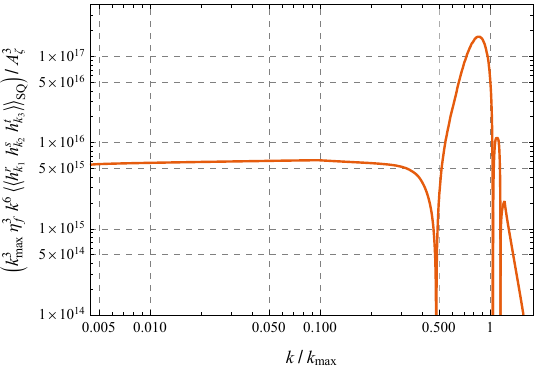}
    \caption{Schematic of the bispectrum for the RRR polarization mode in the squeezed shape. Considering the requirement for the lower limit of the wavenumber $k$, the left boundary of the abscissa is set to $1/225$.}
    \label{fig:SQ}
\end{figure}

 Another shape is the so-called squeezed shape, where the three external momenta $k_i$ satisfy the squeezed limit $k_1 \simeq k_2 \gg k_3$. Without loss of generality \cite{Chen:2010xka}, we set $k_1=k_2=k$, and due to the constraint that the external momenta cannot be smaller than $k_{\min}$, we take $k_3=k_{\max}/225$. Fig.~\ref{fig:SQ} shows the bispectrum curve for the RRR polarization mode under the squeezed limit, with $k_3$ fixed and the other two external momenta $k$ varied. At $k\sim 0.48 k_{\max}$, the bispectrum in the squeezed shape also exhibits a sign flip. The trend in the region $k<k_{\max}$ remains consistent with that of the equilateral shape. The key difference is that in the region $k>k_{\max}$, the squeezed shape does not drop rapidly but instead shows an oscillatory structure. It is worth noting that this oscillation behavior is unique compared to the equilateral shape.

\begin{figure}[!htb]
    \centering
        \begin{subfigure}[b]{0.475\textwidth}
        \includegraphics[width=\textwidth]{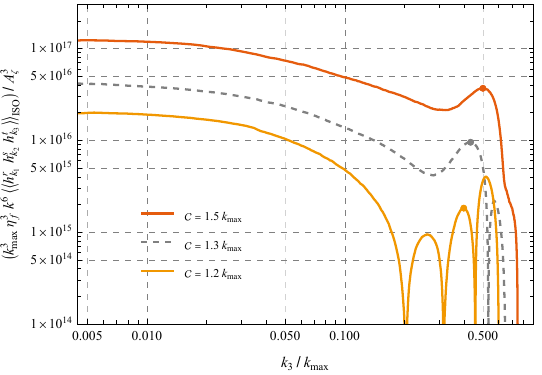}
        \caption{$\mathcal{C}>\mathcal{C}_{\rm c}$}
        \label{fig:ISO+}
     \end{subfigure}
     \begin{subfigure}[b]{0.475\textwidth}
        \includegraphics[width=\textwidth]{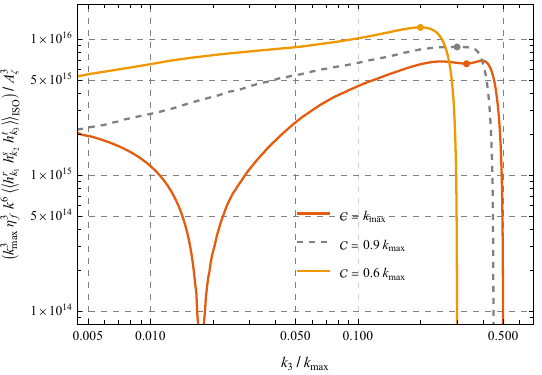}
        \caption{$\mathcal{C}<\mathcal{C}_{\rm c}$}
        \label{fig:ISO-}
     \end{subfigure}
    \caption{The bispectrum with different $\mathcal{C}$, where $C_{\rm c}=1.05k_{\max}$. The positions of the equilateral shape on the bispectrum curve is represented by the dots in the figure. The left boundary is set to $k=k_{\max}/225$, which corresponds to the squeezed shape. }
    \label{fig:ISO}
\end{figure}

To fully characterize the bispectrum's dependence on the shape, we consider the case that fixing the perimeter $\mathcal{C} \equiv k_1+k_2+k_3$ of the triangle formed by the three external momenta. We still assume an isosceles shape where $k_1=k_2=k$ and focus only on the RRR polarization. Before numerical calculations, we first perform a simple analysis of the bispectrum's shape dependence by combining Figs. \ref{fig:EQ-shape} and \ref{fig:SQ}. The equilateral and squeezed shapes exhibit their first sign flip at $k \sim 0.39 k_{\max }$ and $k \sim 0.48 k_{\max }$, respectively, corresponding to $\mathcal{C} \simeq 1.17 k_{\max }$ and $\mathcal{C} \simeq 0.96 k_{\max}$, which implies that near $\mathcal{C}\sim k_{\max}$, their should be a critical behavior. We then plot the bispectrum as a function of $k_3$ for several fixed values of $\mathcal{C}$ in Fig.~\ref{fig:ISO}. The left boundary of the horizontal axis is set to be the squeezed limit, where we choose $k_{\min}=k_{\max}/225$. The dots in each curve represent the equilateral shape. By comparing the trends of different curves, we find that the bispectrum shows a strong dependence on the scale. Meanwhile, as the perimeter $\mathcal{C}$ decreases, the squeezed shape is suppressed relative to the equilateral shape until a critical value $\mathcal{C}_{\mathrm{c}}$, beyond which the bispectrum at the equilateral shape exceeds that of the squeezed shape.

Fig.~\ref{fig:ISO+} corresponds to $\mathcal{C}>\mathcal{C}_{\mathrm{c}}$. Here in the transition model from MD to the RD era, unlike the case of the pure RD case, where the global maximum is located in the equilateral shape, the global maximum occurs in the squeezed shape. We find that as $\mathcal{C}$ decreases, the global maximum of the bispectrum moves downward, which is consistent with the scale dependence of the equilateral shape shown in Fig.~\ref{fig:EQ-shape}. It should be noted that when $\mathcal{C}$ approaches $1.17 k_{\max}$ from the right, two sign flip points simultaneously appear on the left side of the point that represents the equilateral shape, which is a consequence of the requirement for both the squeezed and the equilateral shapes to remain positive. We now turn to Fig.~\ref{fig:ISO-}, that is, $\mathcal{C}<\mathcal{C}_{\mathrm{c}}$ situation, where the equilateral shape is significantly higher than the squeezed shape. The squeezed shape in the curve that represents $\mathcal{C}=k_{\max}$ is positive. We can see that the equilateral shape now is a local minimum but still higher than the squeezed shape. However, when $\mathcal{C}$ decreases to $\mathcal{C}=0.9 k_{\max}$, the sign of the squeezed shape flips from positive to negative, causing the disappearance of all sign flip points. Here, the equilateral shape becomes the global maximum. Then if we increase $\mathcal{C}$ to $\mathcal{C}=1.17 k_{\max}$, the second sign flip point of the yellow curve in Fig.~\ref{fig:ISO+} vanishes, and the equilateral shape transitions from a global maximum to a local minimum.

\begin{figure}[!htb]
    \centering
    \includegraphics[width=\linewidth]{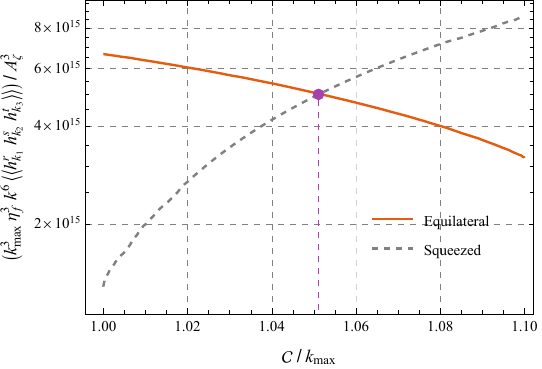}
    \caption{The dependence of the bispectrum on the perimeter $\mathcal{C}$.  The purple dots represent the intersection points of the two curves, corresponding to the case where the bispectrum of the squeezed shape equals that of the equilateral shape.}
    \label{fig:critical}
\end{figure}

To show the scale dependence of the shape in the tensor bispectrum, we plot the dependence of the bispectrum on $\mathcal{C}$ for both the equilateral and squeezed shapes in Fig. \ref{fig:critical}, from which we numerically obtain $\mathcal{C}_{\mathrm{c}} \sim 1.05 k_{\max}$.

This critical behavior of the bispectrum is consistent with physical expectations. As shown in Ref.\ \cite{Inomata:2019ivs}, the power spectrum exhibits a peak at $k \sim k_{\max}$. For large values of $\mathcal{C}$, the two external momentum modes in the squeezed configuration, with $k_1 = k_2 = k \simeq 0.5\mathcal{C}$, lie near the peak of the power spectrum curve, which gives a significant enhancement to the bispectrum in the squeezed configuration. For small values of $\mathcal{C}$, however, the wave numbers $k$ of all modes are small, and these modes tend to enter the horizon near the RD era. This will lead to a similar result than the bispectrum of GWs in pure RD era, where the bispectrum is expected to attain its maximum in the equilateral configuration.

\section{Conclusion and discussion}
\label{se:Conclusion}

In this paper, we for the first time calculate the tensor bispectrum originating from the poltergeist mechanism, which assumes that the Universe goes through a prolonged eMD era and suddenly goes back to the RD era before ucleosynthesis. This scenario leads to a significant resonant amplification of the two-point correlation of tensor perturbations(the power spectrum of GWs) and we also find the same amplification in the tensor bispectrum.

The most interesting part in our results is the configuration dependence of the bispectrum, which has shown a characteristic behavior. As the scale changes, the position of the maximum point will change accordingly. Sometimes, it gives a squeezed configuration, and in other cases, it gives an equilateral configuration. This result is worth noticing because this behavior is different from the non-Gaussian equilateral configuration of GWs generated in the pure RD era and also different from the squeezed configuration bispectrum generated in the inflationary era, where the maximum of the bispectrum is invariably located in the equilateral shape and squeezed shape, respectively. Such complex scaling and configuration dependence provides us with rich information to distinguish different models.



\section*{Acknowledgement}
This work is supported in part by the National Key Research and Development Program of China No. 2020YFC2201501, in part by the National Natural Science Foundation of China No. 12475067 and No. 12235019.

\bibliographystyle{apsrev4-2}
\bibliography{citation}
	

\end{document}